\documentclass[pre,twocolumn,showpacs,preprintnumbers,amsmath,amssymb,floatfix]{revtex4}

\usepackage{graphicx}
\usepackage{dcolumn}
\usepackage{bm}
\usepackage{enumerate}

\usepackage[normalem]{ulem}  
\usepackage[dvips]{color}

\begin{document}


\title{A dynamical model for competing  opinions}

\author{S.R. Souza$^{1,2}$}
\email{srsouza@if.ufrj.br}
\author{S. Gon\c calves$^2$}
\email{sgoncalves@if.ufrgs.br}
\affiliation{$^1$Instituto de F\'\i sica, Universidade Federal do Rio de Janeiro,
Cidade Universit\'aria, \\CP 68528, 21941-972, Rio de Janeiro, Brazil}
\affiliation{$^2$Instituto de F\'\i sica, Universidade Federal do Rio Grande do Sul,\\
Av. Bento Gon\c calves 9500, CP 15051, 91501-970, Porto Alegre, Brazil}

\date{\today}

\begin{abstract}
We propose an opinion model based on agents located at the vertices of a regular lattice.
Each agent has an independent opinion (among an arbitrary, but  fixed, number of choices) and its own degree of conviction.
The latter changes every time it interacts with another agent who has a different opinion.
The dynamics leads to size distributions of clusters (made up of agents which have the same opinion and are located at contiguous spatial positions) which follow a power law, as long as the range of the interaction between the agents is not too short, i.e. the system self-organizes into a critical state.
Short range interactions lead to an exponential cut off in the size distribution and to spatial correlations which cause agents which have the same opinion to be closely grouped.
When the diversity of opinions is  restricted to two, non-consensus dynamic  is observed, with unequal population fractions, whereas consensus is reached if the agents are also allowed to interact  with those which are located far from them.
\end{abstract}

\maketitle

\section{\label{sec:introuduction}Introduction\protect}
\label{sect:introduction}
Statistical mechanics has turned out to be quite successful in modeling many systems whose interaction is, in principle, much more complex than those traditionally studied in physics as, in many cases, the systems are made up of agents which are endowed with intelligence and, therefore, the interaction between them depends on their decisions \cite{reviewSocialDynamics2009,cooperationNowak,socialSystemsPauloMurilo}.
Nevertheless, simple statistical models have been developed for describing social systems \cite{reviewSocialDynamics2009,cooperationNowak,KimIdeas,socialSystemsPauloMurilo}, economy \cite{econophysics1999_2,reviewEconophysics2011_1,econophysics1999_1,fatCat2005}, etc.
Despite the great complexity of such real systems, their main properties can be reproduced by simple models which retain their underlying features.

A great deal of effort has been devoted to developing models for describing the properties of systems made up of agents with competing opinions \cite{KimIdeas,opinionSebastian2005,opinionRoberto2006,reviewSocialDynamics2009,majorityRule2003,opinionStanley2009,NCO2011,localVsSocial2005}.
This is of great relevance as human conflicts very often arise from the simultaneous existence of incompatible opinions in populations.
Different systems, such as hierarchical societies \cite{opinionSebastian2005} or democratic ones \cite{majorityRule2003}, where the agents follow the opinion of the local majority within a group, have, for instance, been investigated.
Most of these models allow the agents to assume only one of  two possible opinions.
Such spin flip models are representative of many real situations which offer only two possibilities and, therefore, are also of great interest \cite{reviewSocialDynamics2009,opinionStanley2009,NCO2011}, besides the similarities with other physical systems.

The evolution of scientific paradigms has been recently modeled in Ref.\ \cite{KimIdeas}. 
The slow decline of old ideas and the quick adoption of new ones are the main characteristics of the model.
Those authors find that the dynamics naturally leads to the replacement of old concepts by a new paradigm, which dominates for a certain period of time, until it is gradually replaced by new ideas.

Inspired by that work, we have developed an opinion model where agents are placed at the vertices of a regular lattice and interact only with those located within a certain range.
Each agent has an opinion and its degree of conviction.
In contrast with many other models, such as those developed in Refs.\ \cite{KimIdeas,majorityRule2003,opinionStanley2009}, for instance,  the interaction between two agents is strictly local, in the sense that it relies only on the agents' properties, i.e. their opinions and convictions.
Their neighborhood has no influence on the interaction.
The latter primarily affects their degree of conviction.
More precisely, the interacting agents' convictions are affected during the interaction.
If the conviction of one of the agents reaches a certain lower bound then its opinion changes to that of the opponent.
We therefore take into account the difficulty in persuading someone who has a strong conviction.
In such a case, it is necessary to change his (her) beliefs prior to the acceptance of the new idea.
Furthermore, we allow the agents to interact with those which are located beyond their first neighbors.

The dynamics of the opinion distribution in populations is then studied in the framework of this model and the remainder of this work is organized as follows.
In Sec.\ \ref{sec:model} we give a detailed description of the model.
The results are presented and discussed in Sec.\ \ref{sec:results} whereas concluding remarks are drawn at Sec.\ \ref{sec:conclusions}.

\section{\label{sec:model}The model\protect}
The system is built on a regular mesh of  $N_x$ horizontal and $N_y$ vertical lines, with periodic boundary conditions.
One agent is placed at each vertex and an opinion $O_i$, among the $N_o$ possible ones, is randomly assigned to the i-th agent, 
 as well as a positive integer $C_i$, which corresponds to its degree of conviction.
 An agent located at vertex $(k,l)$ interacts with any of the neighbors located in vertices $(k',l')\ne (k,l)$, where $k'=k-r,k-r+1,\cdots,k+r-1,k+r$ and $l'=l-r,l-r+1,\cdots,l+r-1,l+r$.
 The range $r$ is one of the model parameters.
 Thus, each agent has $4r(r+1)$ neighbors.
At each step of the dynamics:

\begin{description}
\item{\it a)} An agent $i$ is sampled with probability proportional to $C_i$ and one of its neighbors is randomly selected for interaction, as described below.
By doing so we assume that the agent's activity is related to its conviction.
\item{\it b)} With probability $\alpha$, another agent is randomly selected among the others and its opinion changes to any of the $N_o$ possible ones, including its own.
This procedure represents the replacement of the agent by death or substitution due to departure from its neighborhood.
Its new conviction is selected between 1 and the maximum existing value, in order not to introduce any bias into the system.
\end{description}

In step  (a) above, nothing is done if the agents have the same opinion and one then proceeds to step (b).

Two agents `$i$' and `$j$' may interact only if their opinions are different. In this case, they do it with probability

\begin{equation}
p=\exp[-\lambda (C_{\rm min}/C_{\rm max})^ 2]\;,
\label{eq:pint}
\end{equation}

\noindent
where $C_{\rm min}$ ($C_{\rm max}$) is the minimum (maximum) between $C_i$ and $C_j$, and $\lambda$ is a parameter which is chosen so as to minimize the interaction between agents for which $C_i\approx C_j$.
This is because it is very unlikely that a leader would be influenced by another competing leader (two agents with large and similar values of $C$).
On the other hand, if a leader (large $C$ value) meets an ordinary agent (small $C$ value) who has a different opinion, the latter is very likely to be convinced by the former.
The functional form chosen in this work aims at introducing these features into the model.
If the agents interact, with probability $C_i/(C_i+C_j)$, the conviction $C_i$ is increased by one unit and $C_j$ decreases by $|C_i-C_j|$.
Otherwise, $C_j$ increases by one unit and $C_i$ decreases by $|C_i-C_j|$.
If $C_j\le 0$ ($C_i\le 0$) then its opinion changes to that of agent $i$ ($j$) and $C_j$ ($C_i$) is set to unit.

\begin{figure}[tbh]
\includegraphics[width=8.5cm,angle=0]{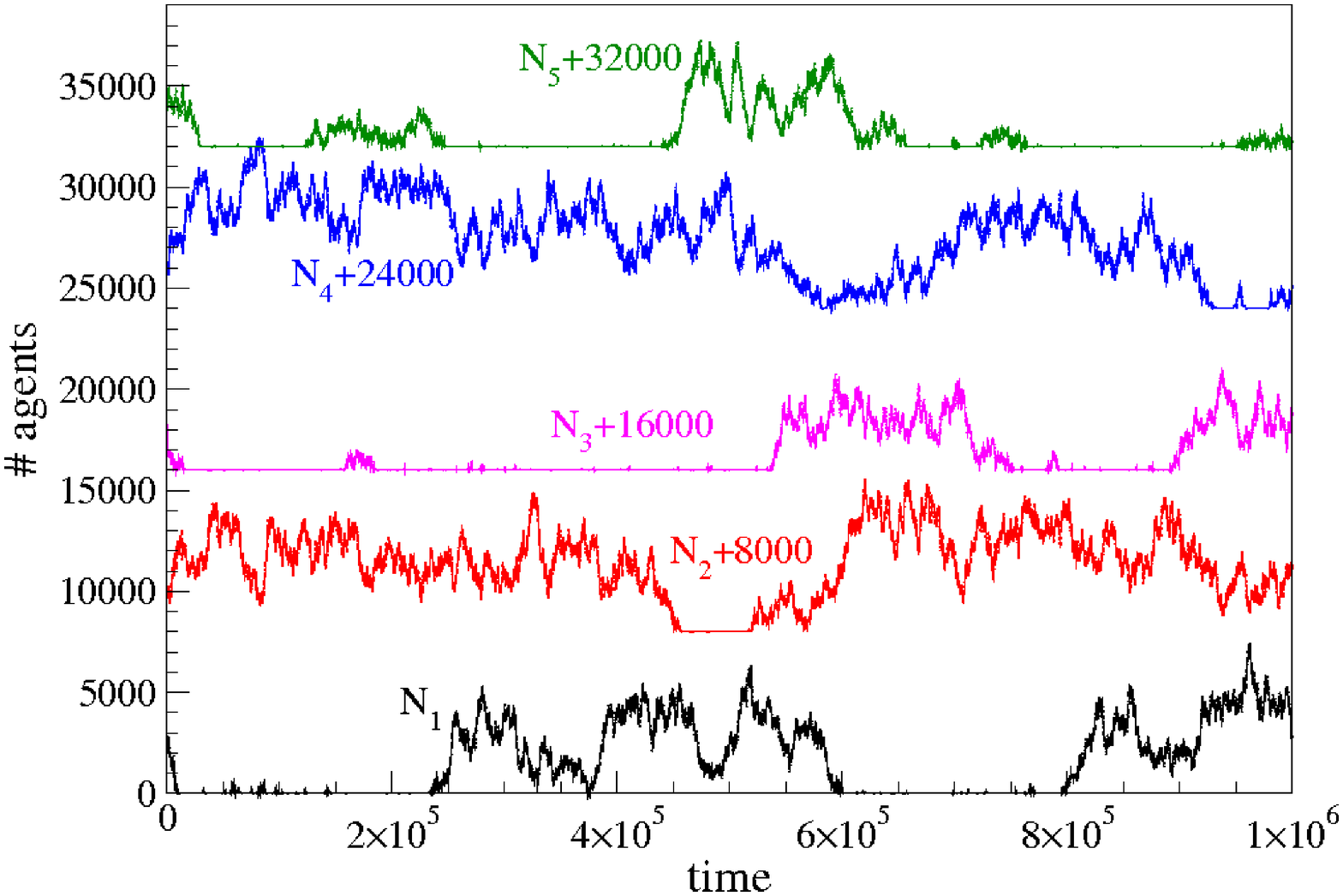}
\caption{\label{fig:popR1} (Color online) Populations of the different opinions as a function of time.
The different groups have been shifted in order to prevent overlap between them.
The curves show the results for $r=1$, $\lambda=1.0$, $N_o=5$, $\alpha=1.0\times 10^ {-6}$, and $N_x=N_y=100$.
For details, see the text.}
\end{figure}

\section{\label{sec:results}Results and discussion\protect}
At the initialization stage, an opinion $1\le O_i\le N_o$ and the conviction degree $1 \le C_i\le 10$ \footnote{The upper bound for $C_i$ used in the initialization does not play a relevant role in the evolution since it is adjusted by the system during the dynamics.} are randomly selected and assigned to the i-th agent.
The system then evolves during, at least, $10^8$ steps.
A full step corresponds to the time interval during which $N_x\times N_y$ intermediate steps, as  explained in (a) and (b) in Sec.\ref{sec:model}, take place.

We start by examining the time evolution of the populations of groups with the same opinion.
The results obtained using $\lambda=1.0$, $r=1$, $N_o=5$, $\alpha=1.0\times 10^ {-6}$, and $N_x=N_y=100$ are shown in Fig.\ \ref{fig:popR1}.
The populations have been shifted in order to prevent overlap between them.
For clarity, we have also restricted the time scale to $10^6$, in spite of having carried out simulations up to much larger times, as just mentioned.
One sees that no opinion dominates the dynamics.
They coexist in different proportions and one notices that, very often, there is one which is much more popular than the others.
Its dominance lasts for a relatively short time and the popular opinion is replaced by another one.
As a matter of fact, only a small number of opinions are effectively disseminated through the system, the others being a small perturbation most of the time.
The most popular opinions are replaced by the unimportant ones, but few opinions dominate the population at the same time.

\begin{figure}[tbh]
\includegraphics[width=8.5cm,angle=0]{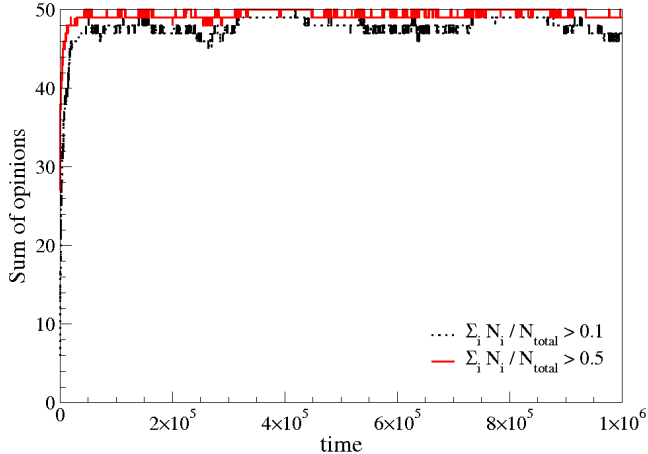}
\caption{\label{fig:sumOps} (Color online) Sum of populations, such that $\sum_iN_i/N_{\rm total} > \epsilon$, $\epsilon=0.1$ and 0.5.
The parameter set is the same used in Fig.\ \ref{fig:popR1}, except for the number of opinions where $N_o=50$ is now employed.
For details, see the text.}
\end{figure}

This conclusion is independent of the number of possible opinions $N_o$ as one sees in Fig.\ \ref{fig:sumOps},
which shows the number of opinions, for which

\begin{equation}
\sum_iN_i/N_{\rm total}>\epsilon
\label{eq:nFrac}
\end{equation}

\noindent
where

\begin{equation}
N_{\rm total}=\sum_{i=1}^{N_o} N_i,
\label{eq:nTotal}
\end{equation}

\noindent
as a function of time, for $\epsilon = 0.1$ and 0.5.
Before carrying out the sum, the indices $i$ are rearranged so that the smallest populations  are selected to calculate the fractions. 
The calculations have been carried out for $N_o=50$.
It is clear from these results that, although the composition may change, only very few opinions are actually adopted by the populations.
Therefore, the use of a fixed number of opinions should not be seen as a limitation of the model as the system naturally eliminates most of the competing opinions and very few of them effectively take part into the dynamics. 
We have checked that this conclusion still holds if one uses different parameters, such as $\lambda=2.0$ or $r=10$.

\begin{figure}[tbh]
\includegraphics[width=8.5cm,angle=0]{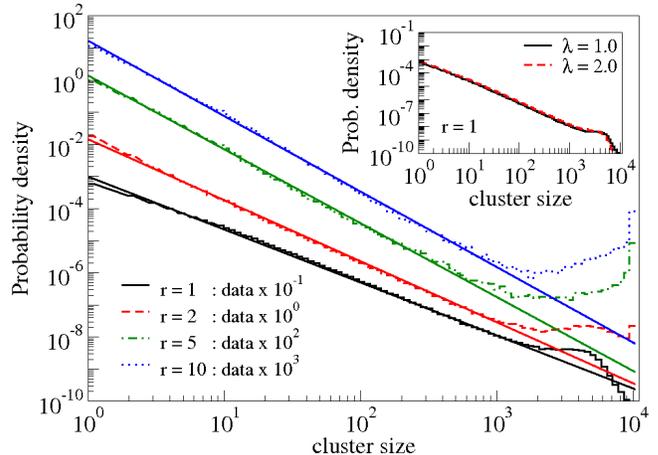}
\caption{\label{fig:sd} (Color online) Size distribution of clusters made up of neighbor agents which have the same opinion.
The parameter set corresponds to that used in Fig.\ \ref{fig:sumOps}.
The power laws are best fit to the results, whose exponents $\beta=1.65,\,1.90,\,2.30,$ and $2.35$ are respectively associated with $r=1,\,2\,,5,$ and $10$.
In the inset, the model predictions for $\lambda=1.0$ are compared to those for $\lambda=2.0$ and $r=1$ in both cases.
For details, see the text.}
\end{figure}

We now turn to the size distribution of clusters made up of neighbors  holding the same opinion.
The system configuration is analyzed at every $10^3$ full steps and the time average is thus performed.
The results are exhibited in Fig. \ref{fig:sd} for different values of the range $r$.
Two agents belong to the same cluster if they have the same opinion and  their grid coordinates  obey $(k,l)-(k',l')=(\pm 1,0),(0,\pm 1)$, or $(\pm 1,\pm 1)$.
The results clearly show that the size distribution is very sensitive to the interaction range $r$.
For cluster sizes up to $10\%$ of the total system, the distribution becomes steeper as $r$ increases, whereas the development of a big cluster, of approximately the size of the total system, becomes more and more pronounced.
Since agents interact only if they do not have the same opinion, the borders between clusters are the regions of strong activity.
For short range interactions, only agents which are located very close to the borders are allowed to interact.
Therefore, for  small values of $r$, one should expect to observe compact groups of agents, who share the same opinion.
This should favor the appearance of medium size clusters.
Indeed, large range values would lead to very diffuse borders and, therefore, to the disappearance of the coherence among the agents which are close to each other.
In the limit of very large $r$ values, the connected agents would pervade the system and the different groups would interpenetrate each other, as they would not be compact.
This would favor the appearance of large clusters during the dynamics, whose contribution to the size distribution may also be noticed in Fig.\ \ref{fig:sd}.
Owing to the strict conservation laws, the multiplicity of small clusters should then decrease and their size distribution would be steeper, as is also seen in Fig.\ \ref{fig:sd}.

\begin{figure}[tbh]
\includegraphics[width=8.5cm,angle=0]{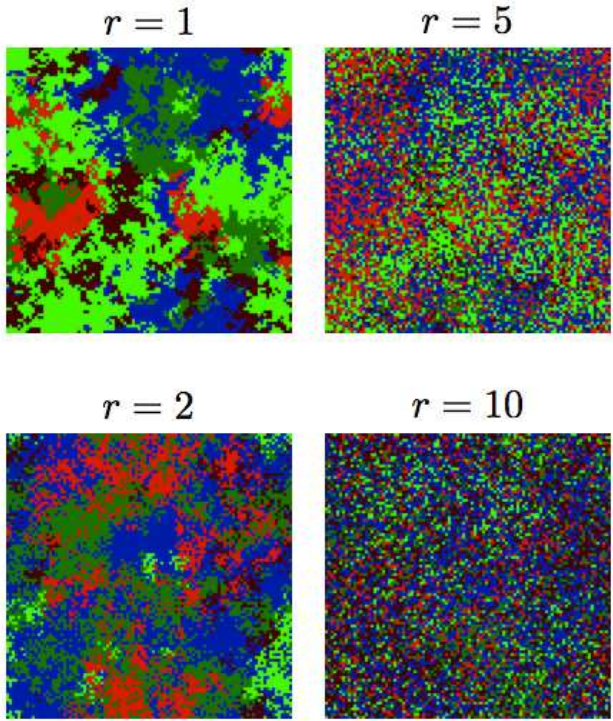}
\caption{\label{fig:clusters} (Color online) Spacial cluster distribution at randomly selected moments of the dynamics. Agents sharing the same opinion are depicted by the same gray values (color). 
The simulation has been carried out using $N_o=5$, $\lambda=1.0$, and $N_x=N_y=100$.
For details, see the text.}
\end{figure}

This qualitative reasoning is confirmed by the results displayed in Fig.\ \ref{fig:clusters}, where the spatial configuration of the clusters is shown at randomly selected moments.
Distinct opinions are represented by different gray values (color online).
One sees that, compact groups are indeed formed for $r=1$, whereas the clusters become more and more spatially diffuse as $r$ increases.
Thus, our model predicts that long range interactions tend to destroy spatial correlations among opinions, when consensus has not been reached and different opinions coexist in the system.

The role played by the parameter $\lambda$ is illustrated in the inset of Fig.\ \ref{fig:sd}, where the cluster size distributions  obtained with $\lambda=1.0$ and $\lambda=2.0$, for $r=1$, are compared.
The effect on the size distribution is small and is more easily noticed at large sizes where one observes a slight suppression of big clusters.
We have also checked that the other observables studied in this work are weakly affected if one changes $\lambda$ in the range $1.0\le \lambda \le 2.0$.
For the sake of simplicity, we adopt $\lambda=1.0$ from here on.

One may also notice in Fig.\ \ref{fig:sd} that the size distribution of clusters, whose size $s$ is smaller than 10\% of the  system size, is fairly accurately approximated by a power law, i.e. $P(s)\propto s^{-\beta}$.
The exponent varies with $r$ and corresponds to $\beta=1.65,\,1.90,\,2.30,$ and $2.35$ for $r=1,\,2,\,5,$ and $10$, respectively.
We have checked that the asymptotic value is reached for $5 < r \le 10$.

These results show that the system self organizes, i.e. the configurations are reached without the need of the external tuning of any parameter, and the power law  suggests the
existence of critical behavior.
The self organized criticality (SOC) has been  discussed in different places \cite{soc1987prl,socPRE1996,socPRA1988,socSneppen1992,BakSOC} and has been observed in many different systems \cite{BakSOC,BakSneppen,Korsnes,iceRussians,socJorgePauloMurilo}.
In our case, this suggests that, on rare occasions,  consensus would spontaneously be reached.
This state should survive for a while, until conflicting opinions are nucleated by the noise described in (b) in Sec.\ \ref{sec:model} and the competition among them would restart and another opinion would dominate, and so on.

\begin{figure}[tbh]
\includegraphics[width=8.5cm,angle=0]{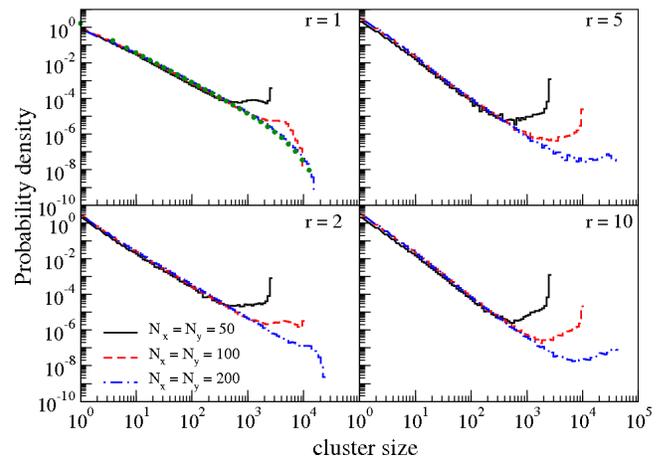}
\caption{\label{fig:sdNxNyR} (Color online) Same as Fig.\ \ref{fig:sd} for different system sizes.
The same parameter set has also been employed in the present simulations.
The full circles in the up left frame correspond to $P(s)\propto \exp(-0.00027s)/s^{1.65}$.
For details, see the text.}
\end{figure}

In order to investigate whether scale invariance is, in fact, present in the dynamics, we show, in Fig.\ \ref{fig:sdNxNyR}, the cluster size distributions  for different $r$ values and system sizes.
The results reveal that, for $r=1$, there seems to be a characteristic scale, since the largest cluster formed during the dynamics does not scale with the system size.
In fact, as shown by the full circles in the up left frame of this figure, the distribution for $r=1$ and $N_x=N_y=200$ is  accurately described by $P(s)\propto \exp(-0.00027s)/s^{1.65}$, which has a clear exponential cut off.
These statements are weakened for $r=2$ and are no longer valid for larger range values.
More precisely, the power law regime extends to larger sizes as the total system size increases for $r\ge 2$.
Actually, there  is a range value, between 2 and 5, for which the power law regime is an adequate description of the size distribution, except for very large clusters, since finite size effects have to be considered for those clusters.
Thus, the dynamics leads to two distinct scenarios.
For small $r$, compact clusters are formed and, occasionally, amalgamate and form a very large one.
Due to the locality of the coalescence process (for small $r$ the interactions take place at the cluster's borders), this happens very rarely since it requires strong spatial correlations.
Then, this process leads to the existence of a size cut off.
On the other hand, when large $r$ values are used, the coalescence extends through much larger areas, due to the spread of the clusters.
It therefore makes it easier for correlations to propagate through the entire system.
Thus, the present model predicts the existence of SOC in systems with competing opinions, if the interaction between the  agents are not restricted to their contiguous neighborhood.

Up to this point, we have not investigated the role played by the parameter $\alpha$, which regulates the frequency with which an agent randomly changes his opinion.
It contributes with noise, which prevents the system from freezing when consensus is reached.
In this sense, the model strongly relies on this parameter to ensure an interesting dynamics.
We have found that it also plays a very important role in determining the cluster size distribution.
We observed that $\alpha=10^{-5}$ still leads to power law regimes for not too large clusters.
However, the invariance of the exponent with the system size shown in Fig.\ \ref{fig:sdNxNyR}, for $r\gtrsim 2$, does not hold in this case.
This means that, although noise is needed by the dynamics,  too much noise destroys the scale invariance, i.e. the agents must keep their opinions for, at least, a short while, in order to preserve spatial correlations.
Since, on the average, $N_x\times N_y \alpha$ agents randomly change their opinions at each step, one sees that there is no unique value of $\alpha$ that would ensure scale invariance for arbitrary system sizes since one may always find a size for which too much noise is added to the system at each time step, destroying  the spatial correlations.
This shortcoming may be avoided by redefining $\alpha$ as the total rate per step, i.e. proportional to $(N_xN_y)^{-1}$, so that the desired amount of noise is introduced into the system during the dynamics, for any system size.

\begin{figure}[tbh]
\includegraphics[width=8.5cm,angle=0]{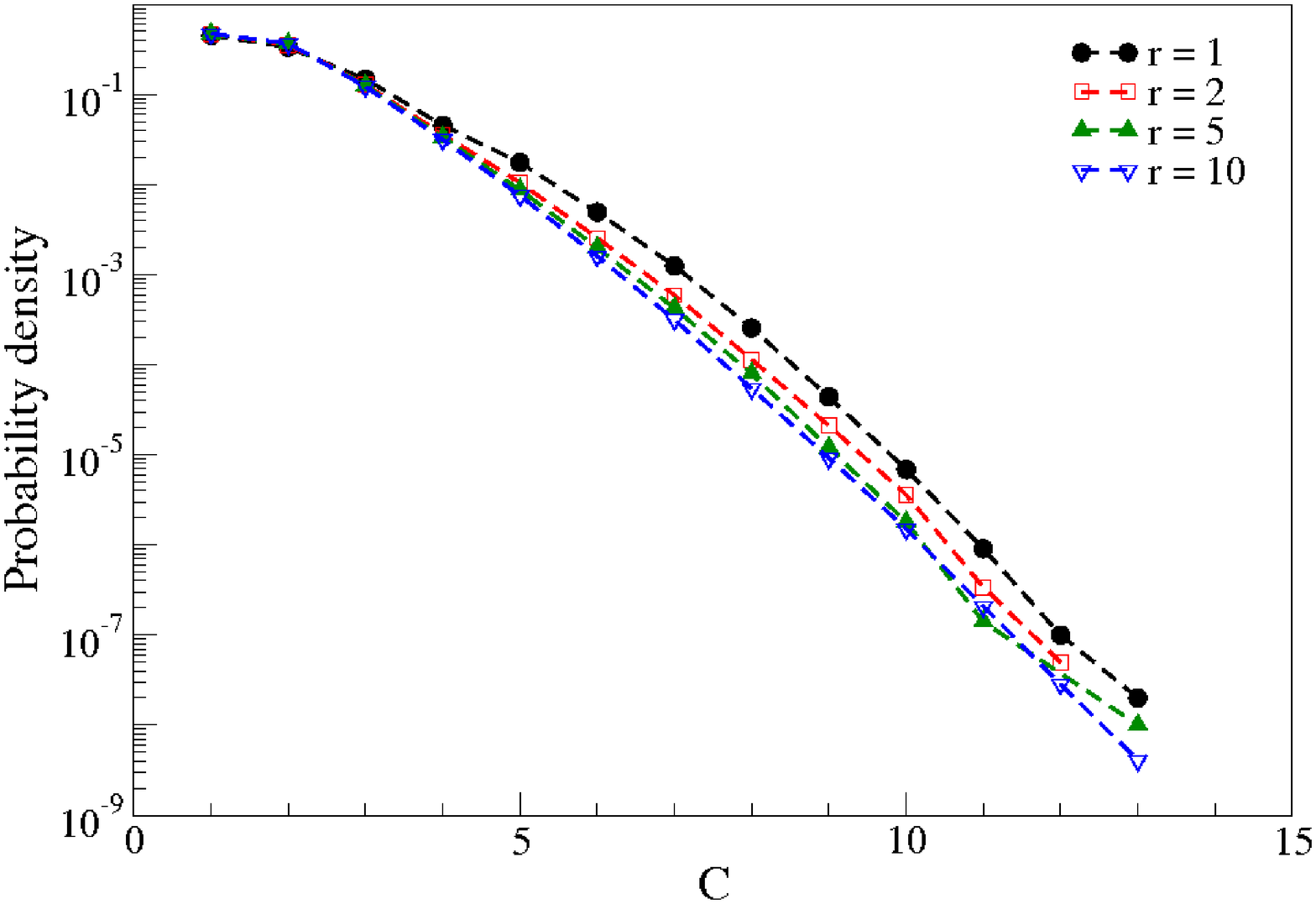}
\caption{\label{fig:cd} (Color online) Time averaged conviction distribution. The parameter set is the same employed in the calculations shown in Fig.\ \ref{fig:sd}.
For details, see the text.}
\end{figure}

Since the agents' convictions play an important role in the dynamics, as it directly influences their resistance to the adoption of new paradigms, we also examine this quantity.
Thus, Fig.\ \ref{fig:cd} displays the time averaged conviction distribution for different values of $r$.
As in real life, most of the agents have a low degree of conviction and the system has very few leaders (large $C$ values), i.e.
the distribution decays exponentially.
As expected, for a given value of $C$, the distribution falls off as $r$ increases, since longer range interaction allows the agents to encounter others with different opinions more often (as is illustrated in Fig. \ref{fig:clusters}, agents with the same opinion tend to be closer for small $r$ values and they do not interact).

\begin{figure}[tbh]
\includegraphics[width=8.5cm,angle=0]{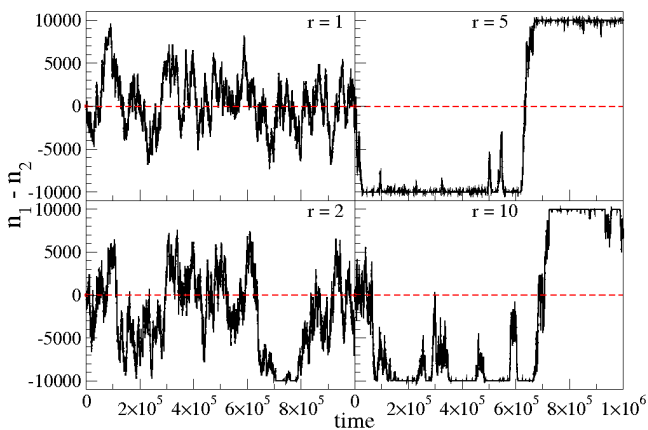}
\caption{\label{fig:diffpop} (Color online) Time evolution of the difference between populations with opinion 1 and 2.
Except for $N_o=2$, the parameter set is the same employed in the calculations shown in Fig.\ \ref{fig:clusters}.
For details, see the text.}
\end{figure}

We finally examine the dynamics when only two opinions are allowed.
The modeling of such systems is of great interest and have been extensively studied \cite{opinionRoberto2006,reviewSocialDynamics2009,opinionStanley2009,NCO2011,localVsSocial2005} since there are many situations in real life where binary choices have to be made \cite{reviewSocialDynamics2009}.
Furthermore, the above results, associated with Fig.\ \ref{fig:sumOps}, also suggest that this scenario should retain most of the properties of real systems.
The results of the model simulation obtained for $N_x=N_y=100$ are displayed in Fig.\ \ref{fig:diffpop}, which exhibits the difference between the populations with opinion 1 ($n_1$) and opinion 2 ($n_2$).
As in Refs.\ \cite{opinionStanley2009,NCO2011}, where a non-consensus opinion model has been proposed and studied, our model allows for the dynamic coexistence of the two conflicting opinions with unequal population fractions.
This is seen in the left panels of this figure which show the results for short range interactions  between neighbors, i.e. $r=1$ and $2$.
This is in agreement with Ref.\ \cite{opinionStanley2009} which considers the interaction between the closest neighbors.
For larger $r$ values, one sees that consensus is reached and it lasts for a long period.
We have followed the dynamics during much longer time scales and confirmed this feature.
Owing to the noise introduced by the random change of the agents' opinions, regulated by the parameter $\alpha$, the status quo does not last forever and, after being the overwhelmingly dominating opinion for a long period, its replacement occurs very quickly and the other opinion becomes the consensus, and so on.
Therefore, our model predicts that a transition from non-consensus to consensus occurs as the interaction among the agents changes from short to long range.

\section{\label{sec:conclusions}Concluding remarks\protect}
We have developed a model for the dynamics of  competing opinions, which is based on the agents' degree of conviction and on the range of the interaction between  them.
It predicts that, even when many different opinions are allowed, only very few of them are really in use by the agents during the dynamics.
This is, in fact, observed in real life when, for instance, at the beginning of an election process, many candidates running for a political office start with not too different opportunities but, after a while, very few dominate the voters' preferences.
The model also predicts that the size distribution of clusters, made up of agents which are located in contiguous spatial positions and share the same opinion, follows a power law.
That distribution is reached independently of the initial conditions, i. e. the dynamics leads to SOC \cite{soc1987prl,socPRE1996,socPRA1988,socSneppen1992,BakSOC}, as long as the interaction between the agents is not restricted to too close neighbors.
When only two opinions are allowed, the model leads to non-consensus dynamics, which qualitatively agrees with the non-consensus model proposed in Ref.\ \cite{opinionStanley2009}.
On the other hand, if the agents also interact with those who are located relatively far from them, consensus is quickly reached and it lasts for a long time.
The dominating opinion is occasionally replaced, but there is consensus almost all the time.
Our model then provides a means to simulate many of the properties of real systems by changing a parameter which has a very simple interpretation on physical systems, i. e. the range of the interaction between the agents.
It also contrast with other models as the interaction between agents affects their conviction in first place and their opinions change only after their paradigms have been corroded.

\begin{acknowledgments}
We would like to acknowledge CNPq,  FAPERJ BBP grant, CNPq-PROSUL, FAPERGS,  the joint PRONEX initiatives of CNPq/FAPERJ under
Contract No.\ 26-111.443/2010 and CNPq/FAPERGS , for partial financial support.
\end{acknowledgments}

\bibliography{manuscript}

\begin{thebibliography}{10}%
\makeatletter
\providecommand \@ifxundefined [1]{%
 \ifx #1\undefined \expandafter \@firstoftwo
 \else \expandafter \@secondoftwo
\fi
}%
\providecommand \@ifnum [1]{%
 \ifnum #1\expandafter \@firstoftwo
 \else \expandafter \@secondoftwo
\fi
}%
\providecommand \enquote [1]{``#1''}%
\providecommand \bibnamefont  [1]{#1}%
\providecommand \bibfnamefont [1]{#1}%
\providecommand \citenamefont [1]{#1}%
\providecommand\href[0]{\@sanitize\@href}%
\providecommand\@href[1]{\endgroup\@@startlink{#1}\endgroup\@@href}%
\providecommand\@@href[1]{#1\@@endlink}%
\providecommand \@sanitize [0]{\begingroup\catcode`\&12\catcode`\#12\relax}%
\@ifxundefined \pdfoutput {\@firstoftwo}{%
 \@ifnum{\z@=\pdfoutput}{\@firstoftwo}{\@secondoftwo}%
}{%
 \providecommand\@@startlink[1]{\leavevmode\special{html:<a href="#1">}}%
 \providecommand\@@endlink[0]{\special{html:</a>}}%
}{%
 \providecommand\@@startlink[1]{%
  \leavevmode
  \pdfstartlink
   attr{/Border[0 0 1 ]/H/I/C[0 1 1]}%
   user{/Subtype/Link/A<</Type/Action/S/URI/URI(#1)>>}%
  \relax
 }%
 \providecommand\@@endlink[0]{\pdfendlink}%
}%
\providecommand \url  [0]{\begingroup\@sanitize \@url }%
\providecommand \@url [1]{\endgroup\@href {#1}{\urlprefix}}%
\providecommand \urlprefix [0]{URL }%
\providecommand \Eprint[0]{\href }%
\@ifxundefined \urlstyle {%
  \providecommand \doi [1]{doi:\discretionary{}{}{}#1}%
}{%
  \providecommand \doi [0]{doi:\discretionary{}{}{}\begingroup
  \urlstyle{rm}\Url }%
}%
\providecommand \doibase [0]{http://dx.doi.org/}%
\providecommand \Doi[1]{\href{\doibase#1}}%
\providecommand \bibAnnote [3]{%
  \BibitemShut{#1}%
  \begin{quotation}\noindent
    \textsc{Key:}\ #2\\\textsc{Annotation:}\ #3%
  \end{quotation}%
}%
\providecommand \bibAnnoteFile [2]{%
  \IfFileExists{#2}{\bibAnnote {#1} {#2} {\input{#2}}}{}%
}%
\providecommand \typeout [0]{\immediate \write \m@ne }%
\providecommand \selectlanguage [0]{\@gobble}%
\providecommand \bibinfo [0]{\@secondoftwo}%
\providecommand \bibfield [0]{\@secondoftwo}%
\providecommand \translation [1]{[#1]}%
\providecommand \BibitemOpen[0]{}%
\providecommand \bibitemStop [0]{}%
\providecommand \bibitemNoStop [0]{.\EOS\space}%
\providecommand \EOS [0]{\spacefactor3000\relax}%
\providecommand \BibitemShut [1]{\csname bibitem#1\endcsname}%
\bibitem{reviewSocialDynamics2009}%
  \BibitemOpen
  \bibfield{author}{%
  \bibinfo {author} {\bibfnamefont{C.}~\bibnamefont{Castellano}}, \bibinfo
  {author} {\bibfnamefont{S.}~\bibnamefont{Fortunato}},\ and\ \bibinfo {author}
  {\bibfnamefont{V.}~\bibnamefont{Loreto}},\ }%
  \bibfield{journal}{%
  \Doi{10.1103/RevModPhys.81.591}{\bibinfo {journal} {Rev. Mod. Phys.}}\ }%
  \textbf{\bibinfo {volume} {81}},\ \bibinfo {pages} {591} (\bibinfo {year}
  {2009})%
  \bibAnnoteFile{NoStop}{reviewSocialDynamics2009}%
\bibitem{cooperationNowak}%
  \BibitemOpen
  \bibfield{author}{%
  \bibinfo {author} {\bibfnamefont{H.}~\bibnamefont{Ohtsuki}}, \bibinfo
  {author} {\bibfnamefont{C.}~\bibnamefont{Hauert}}, \bibinfo {author}
  {\bibfnamefont{L.}~\bibnamefont{E.}},\ and\ \bibinfo {author}
  {\bibfnamefont{M.~A.}\ \bibnamefont{Nowak}},\ }%
  \bibfield{journal}{%
  \Doi{0.1038/nature04605}{\bibinfo {journal} {nature}}\ }%
  \textbf{\bibinfo {volume} {441}},\ \bibinfo {pages} {502} (\bibinfo {year}
  {2006})%
  \bibAnnoteFile{NoStop}{cooperationNowak}%
\bibitem{socialSystemsPauloMurilo}%
  \BibitemOpen
  \bibfield{author}{%
  \bibinfo {author} {\bibfnamefont{D.~S.}\ \bibnamefont{S.~{\protect Moss de
  Oliveira}}, \bibfnamefont{P.~M. C.~{\protect de Oliveira}}},\ }%
  \emph{\bibinfo {title} {Evolution, Money War and Computers}}\ (\bibinfo
  {publisher} {Teubner, Sttutgart-Leipzig},\ \bibinfo {year} {1999})\ ISBN
  \bibinfo {isbn} {3-519-00279-5}%
  \bibAnnoteFile{NoStop}{socialSystemsPauloMurilo}%
\bibitem{KimIdeas}%
  \BibitemOpen
  \bibfield{author}{%
  \bibinfo {author} {\bibfnamefont{S.}~\bibnamefont{Bornholdt}}, \bibinfo
  {author} {\bibfnamefont{M.~H.}\ \bibnamefont{Jensen}},\ and\ \bibinfo
  {author} {\bibfnamefont{K.}~\bibnamefont{Sneppen}},\ }%
  \bibfield{journal}{%
  \Doi{10.1103/PhysRevLett.106.058701}{\bibinfo {journal} {Phys. Rev. Lett.}}\
  }%
  \textbf{\bibinfo {volume} {106}},\ \bibinfo {pages} {058701} (\bibinfo {year}
  {2011})%
  \bibAnnoteFile{NoStop}{KimIdeas}%
\bibitem{econophysics1999_2}%
  \BibitemOpen
  \bibfield{author}{%
  \bibinfo {author} {\bibfnamefont{L.}~\bibnamefont{Laloux}}, \bibinfo {author}
  {\bibfnamefont{P.}~\bibnamefont{Cizeau}}, \bibinfo {author}
  {\bibfnamefont{J.-P.}\ \bibnamefont{Bouchaud}},\ and\ \bibinfo {author}
  {\bibfnamefont{M.}~\bibnamefont{Potters}},\ }%
  \bibfield{journal}{%
  \Doi{10.1103/PhysRevLett.83.1467}{\bibinfo {journal} {Phys. Rev. Lett.}}\ }%
  \textbf{\bibinfo {volume} {83}},\ \bibinfo {pages} {1467} (\bibinfo {year}
  {1999})%
  \bibAnnoteFile{NoStop}{econophysics1999_2}%
\bibitem{reviewEconophysics2011_1}%
  \BibitemOpen
  \bibfield{author}{%
  \bibinfo {author} {\bibfnamefont{A.}~\bibnamefont{Charkraborti}}, \bibinfo
  {author} {\bibfnamefont{I.~M.}\ \bibnamefont{Toke}}, \bibinfo {author}
  {\bibfnamefont{M.}~\bibnamefont{Patriarca}},\ and\ \bibinfo {author}
  {\bibfnamefont{F.}~\bibnamefont{Abergel}},\ }%
  \bibfield{journal}{%
  \Doi{10.1080/14697688.2010.539248}{\bibinfo {journal} {Quantitative
  Finance}}\ }%
  \textbf{\bibinfo {volume} {11}},\ \bibinfo {pages} {911} (\bibinfo {year}
  {2011})%
  \bibAnnoteFile{NoStop}{reviewEconophysics2011_1}%
\bibitem{econophysics1999_1}%
  \BibitemOpen
  \bibfield{author}{%
  \bibinfo {author} {\bibfnamefont{L.}~\bibnamefont{Kador}},\ }%
  \bibfield{journal}{%
  \Doi{10.1103/PhysRevE.60.1441}{\bibinfo {journal} {Phys. Rev. E}}\ }%
  \textbf{\bibinfo {volume} {60}},\ \bibinfo {pages} {1441} (\bibinfo {year}
  {1999})%
  \bibAnnoteFile{NoStop}{econophysics1999_1}%
\bibitem{fatCat2005}%
  \BibitemOpen
  \bibfield{author}{%
  \bibinfo {author} {\bibfnamefont{R.}~\bibnamefont{Donangelo}}, \bibinfo
  {author} {\bibfnamefont{A.}~\bibnamefont{Hansen}}, \bibinfo {author}
  {\bibfnamefont{K.}~\bibnamefont{Sneppen}},\ and\ \bibinfo {author}
  {\bibfnamefont{S.~R.}\ \bibnamefont{Souza}},\ }%
  \bibfield{journal}{%
  \Doi{10.1016/j.physa.2004.09.046}{\bibinfo {journal} {Physica A}}\ }%
  \textbf{\bibinfo {volume} {348}},\ \bibinfo {pages} {496} (\bibinfo {year}
  {2005})%
  \bibAnnoteFile{NoStop}{fatCat2005}%
\bibitem{opinionSebastian2005}%
  \BibitemOpen
  \bibfield{author}{%
  \bibinfo {author} {\bibfnamefont{M.~F.}\ \bibnamefont{Laguna}}, \bibinfo
  {author} {\bibfnamefont{S.~R.}\ \bibnamefont{Gusman}}, \bibinfo {author}
  {\bibfnamefont{G.}~\bibnamefont{Abramson}}, \bibinfo {author}
  {\bibfnamefont{S.}~\bibnamefont{Gonçalves}},\ and\ \bibinfo {author}
  {\bibfnamefont{J.~R.}\ \bibnamefont{Iglesias}},\ }%
  \bibfield{journal}{%
  \Doi{10.1016/j.physa.2004.11.064}{\bibinfo {journal} {Physica A}}\ }%
  \textbf{\bibinfo {volume} {351}},\ \bibinfo {pages} {580} (\bibinfo {year}
  {2005})%
  \bibAnnoteFile{NoStop}{opinionSebastian2005}%
\bibitem{opinionRoberto2006}%
  \BibitemOpen
  \bibfield{author}{%
  \bibinfo {author} {\bibfnamefont{M.~S.}\ \bibnamefont{{\protect de la
  Lama}}}, \bibinfo {author} {\bibfnamefont{I.~G.}\ \bibnamefont{Szendro}},
  \bibinfo {author} {\bibfnamefont{J.~R.}\ \bibnamefont{Iglesias}},\ and\
  \bibinfo {author} {\bibfnamefont{H.~S.}\ \bibnamefont{Wio}},\ }%
  \bibfield{journal}{%
  \bibinfo {journal} {Eur. Phys. J. B.}}%
   (\bibinfo {year} {2006})%
  \bibAnnoteFile{NoStop}{opinionRoberto2006}%
\bibitem{majorityRule2003}%
  \BibitemOpen
  \bibfield{author}{%
  \bibinfo {author} {\bibfnamefont{P.~L.}\ \bibnamefont{Krapivsky}}\ and\
  \bibinfo {author} {\bibfnamefont{S.}~\bibnamefont{Redner}},\ }%
  \bibfield{journal}{%
  \bibinfo {journal} {Phys. Rev. Lett.}\ }%
  \textbf{\bibinfo {volume} {90}},\ \bibinfo {pages} {238701} (\bibinfo {year}
  {2003})%
  \bibAnnoteFile{NoStop}{majorityRule2003}%
\bibitem{opinionStanley2009}%
  \BibitemOpen
  \bibfield{author}{%
  \bibinfo {author} {\bibfnamefont{J.}~\bibnamefont{Shao}}, \bibinfo {author}
  {\bibfnamefont{S.}~\bibnamefont{Havlin}},\ and\ \bibinfo {author}
  {\bibfnamefont{H.~E.}\ \bibnamefont{Stanley}},\ }%
  \bibfield{journal}{%
  \Doi{10.1103/PhysRevLett.103.018701}{\bibinfo {journal} {Phys. Rev. Lett.}}\
  }%
  \textbf{\bibinfo {volume} {103}},\ \bibinfo {pages} {018701} (\bibinfo {year}
  {2009})%
  \bibAnnoteFile{NoStop}{opinionStanley2009}%
\bibitem{NCO2011}%
  \BibitemOpen
  \bibfield{author}{%
  \bibinfo {author} {\bibfnamefont{D.}~\bibnamefont{\protect~ben Avraham}},\ }%
  \bibfield{journal}{%
  \Doi{10.1103/PhysRevE.83.050101}{\bibinfo {journal} {Phys. Rev. E}}\ }%
  \textbf{\bibinfo {volume} {83}},\ \bibinfo {pages} {050101} (\bibinfo {year}
  {2011})%
  \bibAnnoteFile{NoStop}{NCO2011}%
\bibitem{localVsSocial2005}%
  \BibitemOpen
  \bibfield{author}{%
  \bibinfo {author} {\bibfnamefont{S.}~\bibnamefont{Galam}},\ }%
  \bibfield{journal}{%
  \Doi{10.1209/epl/i2004-10526-5}{\bibinfo {journal} {Europhys. Lett.}}\ }%
  \textbf{\bibinfo {volume} {70}},\ \bibinfo {pages} {705} (\bibinfo {year}
  {2005})%
  \bibAnnoteFile{NoStop}{localVsSocial2005}%
\bibitem{Note1}%
  \BibitemOpen
  \bibinfo {note} {The upper bound for $C_i$ used in the initialization does
  not play a relevant role in the evolution since it is adjusted by the system
  during the dynamics.}%
  \bibAnnoteFile{Stop}{Note1}%
\bibitem{soc1987prl}%
  \BibitemOpen
  \bibfield{author}{%
  \bibinfo {author} {\bibfnamefont{P.}~\bibnamefont{Bak}}, \bibinfo {author}
  {\bibfnamefont{C.}~\bibnamefont{Tang}},\ and\ \bibinfo {author}
  {\bibfnamefont{K.}~\bibnamefont{Wiesenfeld}},\ }%
  \bibfield{journal}{%
  \Doi{10.1103/PhysRevLett.59.381}{\bibinfo {journal} {Phys. Rev. Lett.}}\ }%
  \textbf{\bibinfo {volume} {59}},\ \bibinfo {pages} {381} (\bibinfo {month}
  {Jul}\ \bibinfo {year} {1987}),\
  \url{http://link.aps.org/doi/10.1103/PhysRevLett.59.381}%
  \bibAnnoteFile{NoStop}{soc1987prl}%
\bibitem{socPRE1996}%
  \BibitemOpen
  \bibfield{author}{%
  \bibinfo {author} {\bibfnamefont{M.}~\bibnamefont{Paczuski}}, \bibinfo
  {author} {\bibfnamefont{S.}~\bibnamefont{Maslov}},\ and\ \bibinfo {author}
  {\bibfnamefont{P.}~\bibnamefont{Bak}},\ }%
  \bibfield{journal}{%
  \Doi{10.1103/PhysRevE.53.414}{\bibinfo {journal} {Phys. Rev. E}}\ }%
  \textbf{\bibinfo {volume} {53}},\ \bibinfo {pages} {414} (\bibinfo {month}
  {Jan}\ \bibinfo {year} {1996}),\
  \url{http://link.aps.org/doi/10.1103/PhysRevE.53.414}%
  \bibAnnoteFile{NoStop}{socPRE1996}%
\bibitem{socPRA1988}%
  \BibitemOpen
  \bibfield{author}{%
  \bibinfo {author} {\bibfnamefont{P.}~\bibnamefont{Bak}}, \bibinfo {author}
  {\bibfnamefont{C.}~\bibnamefont{Tang}},\ and\ \bibinfo {author}
  {\bibfnamefont{K.}~\bibnamefont{Wiesenfeld}},\ }%
  \bibfield{journal}{%
  \Doi{10.1103/PhysRevA.38.364}{\bibinfo {journal} {Phys. Rev. A}}\ }%
  \textbf{\bibinfo {volume} {38}},\ \bibinfo {pages} {364} (\bibinfo {month}
  {Jul}\ \bibinfo {year} {1988}),\
  \url{http://link.aps.org/doi/10.1103/PhysRevA.38.364}%
  \bibAnnoteFile{NoStop}{socPRA1988}%
\bibitem{socSneppen1992}%
  \BibitemOpen
  \bibfield{author}{%
  \bibinfo {author} {\bibfnamefont{K.}~\bibnamefont{Sneppen}},\ }%
  \bibfield{journal}{%
  \Doi{10.1103/PhysRevLett.69.3539}{\bibinfo {journal} {Phys. Rev. Lett.}}\ }%
  \textbf{\bibinfo {volume} {69}},\ \bibinfo {pages} {3539} (\bibinfo {year}
  {1992}),\ \url{http://link.aps.org/doi/10.1103/PhysRevLett.69.3539}%
  \bibAnnoteFile{NoStop}{socSneppen1992}%
\bibitem{BakSOC}%
  \BibitemOpen
  \bibfield{author}{%
  \bibinfo {author} {\bibfnamefont{P.}~\bibnamefont{Bak}},\ }%
  \emph{\bibinfo {title} {How Nature Works: The Science of Self-Organized
  Criticality}}\ (\bibinfo {publisher} {Copernicus, New York},\ \bibinfo {year}
  {1996})\ ISBN \bibinfo {isbn} {ISBN 0-387-94791-4}%
  \bibAnnoteFile{NoStop}{BakSOC}%
\bibitem{BakSneppen}%
  \BibitemOpen
  \bibfield{author}{%
  \bibinfo {author} {\bibfnamefont{P.}~\bibnamefont{Bak}}\ and\ \bibinfo
  {author} {\bibfnamefont{K.}~\bibnamefont{Sneppen}},\ }%
  \bibfield{journal}{%
  \Doi{10.1103/PhysRevLett.71.4083}{\bibinfo {journal} {Phys. Rev. Lett.}}\ }%
  \textbf{\bibinfo {volume} {71}},\ \bibinfo {pages} {4083} (\bibinfo {month}
  {Dec}\ \bibinfo {year} {1993}),\
  \url{http://link.aps.org/doi/10.1103/PhysRevLett.71.4083}%
  \bibAnnoteFile{NoStop}{BakSneppen}%
\bibitem{Korsnes}%
  \BibitemOpen
  \bibfield{author}{%
  \bibinfo {author} {\bibfnamefont{R.}~\bibnamefont{Korsnes}}, \bibinfo
  {author} {\bibfnamefont{S.~R.}\ \bibnamefont{Souza}}, \bibinfo {author}
  {\bibfnamefont{R.}~\bibnamefont{Donangelo}}, \bibinfo {author}
  {\bibfnamefont{A.}~\bibnamefont{Hansen}}, \bibinfo {author}
  {\bibfnamefont{M.}~\bibnamefont{Paczuski}},\ and\ \bibinfo {author}
  {\bibfnamefont{K.}~\bibnamefont{Sneppen}},\ }%
  \bibfield{journal}{%
  \bibinfo {journal} {Physica A}\ }%
  \textbf{\bibinfo {volume} {331}},\ \bibinfo {pages} {291} (\bibinfo {year}
  {2004})%
  \bibAnnoteFile{NoStop}{Korsnes}%
\bibitem{iceRussians}%
  \BibitemOpen
  \bibfield{author}{%
  \bibinfo {author} {\bibfnamefont{A.}~\bibnamefont{Chmel}}, \bibinfo {author}
  {\bibfnamefont{V.~N.}\ \bibnamefont{Smirnov}},\ and\ \bibinfo {author}
  {\bibfnamefont{L.~V.}\ \bibnamefont{Panov}},\ }%
  \bibfield{journal}{%
  \bibinfo {journal} {Ocean Science}\ }%
  \textbf{\bibinfo {volume} {3}},\ \bibinfo {pages} {291} (\bibinfo {year}
  {2007})%
  \bibAnnoteFile{NoStop}{iceRussians}%
\bibitem{socJorgePauloMurilo}%
  \BibitemOpen
  \bibfield{author}{%
  \bibinfo {author} {\bibfnamefont{J.}~\bibnamefont{{\protect S\'a-Martins}}}\
  and\ \bibinfo {author} {\bibfnamefont{P.~M.~C.}\ \bibnamefont{{\protect de
  Oliveira}}},\ }%
  \bibfield{journal}{%
  \bibinfo {journal} {Braz. J. Phys.}\ }%
  \textbf{\bibinfo {volume} {34}},\ \bibinfo {pages} {1077} (\bibinfo {year}
  {2004}),\ \url{http://www.sbfisica.org.br/bjp}%
  \bibAnnoteFile{NoStop}{socJorgePauloMurilo}%
\end{thebibliography}%
\end{document}